\documentclass{llncs}
\usepackage{times}
\usepackage[utf8]{inputenc}
\usepackage{setspace}
\usepackage{etex}
\usepackage{latexsym}
\usepackage{amssymb}
\usepackage{mathrsfs}
\usepackage{xcolor}
\usepackage{pictex}
\usepackage{float}
\usepackage{supertabular}
\usepackage{pict2e}
\usepackage{slashbox}
\usepackage[english]{babel}
\usepackage{graphicx}

\newcommand{\omt}[1]{}

\title{Conauto-2.0: Fast Isomorphism Testing and Automorphism Group Computation}

\author{Jos\'e Luis L\'opez-Presa\inst{1}
\and
Antonio Fern\'andez Anta\inst{2}
\and
Luis N\'u\~nez Chiroque\inst{1}}

\institute{DIATEL, Universidad Polit\'ecnica de Madrid, Madrid, Spain
\and
Institute IMDEA Networks, Madrid, Spain}

\newenvironment{packed_enum}{
\begin{enumerate}
  \setlength{\itemsep}{1pt}
  \setlength{\parskip}{0pt}
  \setlength{\parsep}{0pt}
}{\end{enumerate}}

\newenvironment{packed_itemize}{
\begin{itemize}
  \setlength{\itemsep}{1pt}
  \setlength{\parskip}{0pt}
  \setlength{\parsep}{0pt}
}{\end{itemize}}

\newenvironment{proofof}[1]{\noindent{\bf Proof of #1:}}{\hfill \rule{2mm}{2mm}\\}

\begin{document}

\newtheorem{observation}{Observation}

\newcounter{algline}

\newcommand{\nl}{\\ \>\arabic{algline}\' \stepcounter{algline}}

\newenvironment{algo}[1]{\setcounter{algline}{1}\begin{tabbing} 123\=\=678\=123\=678\=123\=678\=123\=678\=123\=678\=  \kill {\bf #1} \nl}{\end{tabbing}}

\newenvironment{boxfig}[1]{\begin{figure}[!tb]\fbox{\begin{minipage}{0.98\linewidth}
                        \vspace{1em}
                        \makebox[0.025\linewidth]{}
                        \begin{minipage}{0.95\linewidth}
                        #1
                        \end{minipage}
                        \end{minipage}}}{\end{figure}}

\newcommand{\B}{\vspace*{-\smallskipamount}}
\newcommand{\BB}{\vspace*{-\medskipamount}}
\newcommand{\BBB}{\vspace*{-\bigskipamount}}

\floatstyle{ruled}
\newfloat{Algorithm}{!tb}{loa}

\maketitle

\begin{abstract}
In this paper we present an algorithm, called \emph{conauto-2.0}, that can efficiently compute a set of generators of the automorphism group of a graph, and test whether two graphs are isomorphic, finding an isomorphism if they are. This algorithm uses the basic individualization/refinement technique, and is an improved version of the algorithm \emph{conauto}, which has been shown to be very fast for random graphs and several families of hard graphs \cite{DBLP:conf/wea/Lopez-PresaA09}.
In this paper, it is proved that, under some circumstances, it is not only possible to prune the search space (using already found generators of the automorphism group), but also to infer new generators without the need of explicitly finding an automorphism of the graph. This result is especially suited for graphs with regularly connected components, and can be applied in any isomorphism testing and canonical labeling algorithm (that use the individualization/refinement technique) to significantly improve its performance.
Additionally, a dynamic target cell selection function is used to adapt to different graphs.
The resulting algorithm preserves all the nice features of \emph{conauto}, but reduces the time for testing graphs with regularly connected components and other hard graph families.
We run extensive experiments, which show that the most popular algorithms (namely, \emph{nauty} \cite{McK81,nautyP}, \emph{bliss} \cite{DBLP:conf/alenex/JunttilaK07,springerlink:10.1007/978-3-642-19754-3_16}, \emph{Traces} \cite{DBLP:journals/corr/abs-0804-4881}, and \emph{saucy} \cite{DBLP:conf/dac/DargaLSM04,DBLP:conf/sat/KatebiSM10}) are slower than \emph{conauto-2.0}, among others, for the graph families based on components.
\end{abstract}


\section{Introduction}
\label{intoduction}
The Graph Isomorphism problem (GI) has been puzzling computer scientists for several decades, because it could not be shown to have polynomial complexity, but it could not be proven to be NP-complete either. However, testing if two graphs are isomorphic is useful in a number of contexts, ranging from chemistry \cite{Faulon98,TinKlin99} to computer vision \cite{ConteFSV03}. Hence, the design of algorithms and the construction of tools that are able to solve the problem efficiently for a large collection of problem instances has
significant interest.

\paragraph{Related Work}

For the last three decades, \emph{nauty} \cite{McK81,nautyP} has been the most widely used tool for graph isomorphism testing and canonical labeling. However, Miyazaki proved \cite{Miyazaki97} that nauty required exponential time for a family of colored graphs. McKay noted also that nauty would also require exponential time for unions of strongly regular graphs. All this has encouraged researchers to develop new tools, to try to overcome this drawback. Some of them are based, like nauty, on canonical labeling. Examples are \emph{bliss} \cite{DBLP:conf/alenex/JunttilaK07,springerlink:10.1007/978-3-642-19754-3_16}, \emph{Traces} \cite{DBLP:journals/corr/abs-0804-4881}, and \emph{nishe} \cite{Tener_Deo_2008,Tener2009}. Another tool, named \emph{saucy} \cite{DBLP:conf/dac/DargaLSM04,DBLP:conf/sat/KatebiSM10} solves a related problem: computing the automorphism group of a graph. It is specially designed to efficiently process big sparse graphs.

A different way to tackle the GI problem was suggested in \cite{DBLP:conf/wea/Lopez-PresaA09}. The tool developed in that work, called \emph{conauto}, does not generate a canonical labeling of the graphs being tested, but instead looks for similar sequences of vertex partitions. To do so, it uses a limited search for automorphisms on the graphs. This algorithm has good performance in practice for different families of graphs, like for example the graphs of Miyazaki. However, its inability of computing the whole automorphism group restricts the benefit obtained from known automorphisms. That reduced its performance with some families of graphs, like Latin square graphs or the point-line graphs of Desarguesian projective planes. One major contribution of the followup work \cite{jllopez2009} was a way to avoid backtracking when processing union graphs, which could be applied in a more general field. Some ways to improve algorithm conauto were also suggested as open problems; in particular, computing the full automorphism group of the graphs, and recording and using information on non-isomorphisms to prune the search space\footnote{The latest versions of bliss (versions 0.65 and above) \cite{springerlink:10.1007/978-3-642-19754-3_16} apply similar ideas: component recursion, and use of failures to prune the search space. However, bliss-0.72 stops with an internal error when processing some graphs (based on unions of connected components). The authors of bliss have been notified of the problem.}.

\paragraph{Contributions}

In this paper we describe a new algorithm, called \emph{conauto-2.0} that, using the basic approaches of \cite{DBLP:conf/wea/Lopez-PresaA09}, extends the functionality of the algorithm, and significantly improves its performance. The main contribution is a new theorem that is applicable under very simple conditions, and prunes the search for automorphisms. This has a large impact in many classes of graphs, like, for instance, graphs built from connected components. 
This result is directly applicable to any other algorithm that uses the individualization/refinement approach to compute the automorphism group or the canonical labeling of a graph. 

One key issue of every automorphism group computation or canonical labeling algorithm is \emph{cell selection}. In many cases, the target cell chosen for individualization conditions the ability of the algorithm to effectively find the generators of the automorphism group. This is particularly critical in the case of Desarguesian projective planes. In fact, knowing how an algorithm selects the target cell enables the construction of graphs that make an algorithm exponential in time. That was the approach of Miyazaki against nauty. While nauty and bliss use a static cell selector, Traces uses a dynamic cell selector that uses \emph{multi-refinements}. However, although this approach shows especially powerful for non-Desarguesian projective planes, it is too expensive for simpler graphs. In conauto-2.0, a dynamic cell selector is used but, for efficiency, it is not isomorphism invariant and, hence, cannot be used for canonical labeling. However, it is able to choose the best target cell in the case of many different graph families (for example the Desarguesian projective planes, or Miyazaki's graphs).
Finally, two additional features of conauto-2.0 are that, \textbf{in addition to test for isomorphism, it computes the complete automorphism group of a graph} (set of generators, size of the automorphism group, and orbits), and it uses detected non-automorphisms in a similar way to the failures management of bliss \cite{springerlink:10.1007/978-3-642-19754-3_16}.

We have carried out extensive experiments to compare the practical performance of conauto-2.0 with nauty-2.4, bliss-0.35, and bliss-0.72\footnote{Two versions of bliss have been considered because bliss-0.72, although faster than bliss-0.35 in some cases, sometimes crashes, as was mentioned.} to test for isomorphism
with several graph families. (Only a small subset of results has been included here.) Our experiments show  that conauto-2.0 is the fastest for most graph families, and when it is not, its performance is similar to that of the others. We have also evaluated the performance of several programs to compute the automorphism group of several selected graphs. The results show that the best performance is presented by bliss0.72, Traces, and conauto-2.0, without a clear
winner among them.


\paragraph{Structure}
Next section defines the basic notion of sequence of partitions on which conauto-2.0 is based. Then the key features of the algorithm are described. In Section~\ref{s-correctness}, the main theorem is presented. Finally, a practical performance evaluation is done.

\section{Definitions and Notation}
\label{theory}

A {\em directed graph} $G$ is the pair $(V,R)$ where $V$ is a finite non-empty set of vertices and $R$ is a binary
relation.
The elements of $R$ are called {\em arcs}. An arc $(u,v) \in R$ is oriented from $u$ to $v$. 
We use the term {\em graph} to refer to a {\em directed graph}\footnote{An {\em undirected graph} is a graph whose arc set $R$ is symmetrical, i.e. $(u,v) \in R$ iff $(v,u) \in R$.}.
For the sake of clarity we may not be fully formal in the following definitions. More details and formal definitions can be found in \cite{DBLP:conf/wea/Lopez-PresaA09,jllopez2009}.

An \emph{isomorphism} of graphs $G=(V_G,R_G)$ and $H=(V_H,R_H)$ is a bijection between the vertex sets
of then, $f:V_G \longrightarrow V_H$, such that $(v,u) \in R_G$ iff $(f(v),f(u)) \in R_H$.
Graphs $G$ and $H$ are {\em isomorphic}, written $G \simeq H$, if there is some
\emph{isomorphism} of them. An {\em automorphism} of $G$ is an isomorphism of $G$ and itself.

Given a graph $G=(V,R)$, we assume $R$ to be given as an {\em adjacency matrix}
$\mathit{Adj}(G)=M$ with size $|V| \times |V|$ as follows.
\begin{displaymath}
M_{uv} = \left\{ \begin{array}{ll | ll}
        0 & \textrm{if $(u,v) \notin R \land (v,u) \notin R$} \; & \; 1 & \textrm{if $(u,v) \notin R \land (v,u) \in R$}\\
        2 & \textrm{if $(u,v) \in R \land (v,u) \notin R$} \; & \;3 & \textrm{if $(u,v) \in R \land (v,u) \in R$.}
\end{array} \right.
\end{displaymath}
Given a vertex $v\in V$ and a subset $V' \subseteq V$, the \emph{available degree} of $v$ with $V'$ is the 3-tuple $(D_3,D_2,D_1)$ counting the number of vertices
in $V'$ connected to $v$ with the 3 types of adjacencies 3, 2, and 1, respectively. If all the elements of the tuple are zero then, $v$ is disconnected from $V'$. The available degree
is used in our algorithms to order the vertices of $G$.

%


\paragraph{Sequences of partitions}

The conauto \cite{DBLP:conf/wea/Lopez-PresaA09,jllopez2009} algorithms all use the same basic approach to test for isomorphism between graphs $G$ and $H$: they build a sequence of partitions for one of the graphs and try to find another sequence of partitions for the other graph with the same underlying structure (compatible). A sequence of partitions defines an ordering of the vertices of the graph. Then, if compatible sequences of partitions are found, the corresponding orderings yield the isomorphism between $G$ and $H$.


Let us consider a graph $G=(V,R)$.
A \emph{partition} of a set $S \subseteq V$ is a sequence $\mathcal{S} = (S_1, ..., S_r)$ of disjoint
nonempty subsets of $S$ such that $S = \bigcup_{i=1}^r S_i$.  The sets $S_i$ are called
the \emph{cells} of $\mathcal{S}$.
A \emph{sequence of partitions} of $G$ is a list of partitions $\mathcal{S}^0,...,\mathcal{S}^t$. The sequence
starts with the trivial partition $\mathcal{S}^0 = (V)$, and partition $\mathcal{S}^{k+1}$ is obtained
from partition $\mathcal{S}^k$ using some refinement. Each refinement can be a \emph{set refinement} or a \emph{vertex refinement} (or vertex individualization).
In both cases a pivot cell is chosen. In a set refinement, each cell of the initial partition $\mathcal{S}^k$ is divided into smaller cells in $\mathcal{S}^{k+1}$ as a function of the
available degree of each of its vertices with respect to the pivot cell. In the vertex refinement, a vertex $p$ of the pivot cell is chosen, and each
cell of the initial partition is divided into smaller cells as a function of the
adjacency of each of its vertices with respect to $p$.
Each refinement can discard some vertices from the partition, either because they have been used as pivot
vertex in a vertex refinement, or because they have no more links with the rest of vertices in the partition.
The sequence ends when all remaining cells are singleton.
In a sequence of partitions, \emph{level} $k$ refers to $\mathcal{S}^k$ and its associated
parameters (e.g., type of refinement, and pivot cell used).

Consider two graphs $G$ and $H$, and sequences of partitions $\mathcal{S}^0,...,\mathcal{S}^t$ and $\mathcal{T}^0,...,\mathcal{T}^t$ of them.
We say that the sequences are \emph{compatible} if both have the same number of levels $t$, at each level the corresponding partitions are compatible as well (i.e., have the same number of cells, pairwise with the same size and the same available degree between them), the refinements applied at each level $k$ is the same, and the pivot cell is in the same position, and the
final partitions $\mathcal{S}^t$ and $\mathcal{T}^t$ have the same adjacencies between cells.
The following theorem shows that having compatible sequences of partitions is equivalent to being isomorphic.

\begin{theorem}[\cite{DBLP:conf/wea/Lopez-PresaA09}]
\label{iso-iif-seq}
Two graphs $G$ and $H$ are isomorphic iff there are two compatible sequences of partitions $\mathcal{S}^0,...,\mathcal{S}^t$ and $\mathcal{T}^0,...,\mathcal{T}^t$ for graphs $G$ and $H$ respectively.
\end{theorem}

The challenge to derive fast isomorphism testing algorithms based on sequence of partitions is to deal with backtracking in the search for a compatible sequence of partitions.
Backtracking occurs when vertex refinement is used in the original sequence,
and the pivot cell has more than one vertex. Our algorithms always guarantee that this happens only when the partition has no singleton cells 
and it is not possible to refine the partition by means of a set refinement. When this happens at level $k$, we say that we have a \emph{backtracking point} at that level, or that $k$ is a
\emph{backtracking level}.

In order to deal with backtracking points, the conauto algorithms use vertex equivalence and automorphism detection. 
In a graph $G=(V,R)$, two vertices $u, v \in V$ are \emph{equivalent at the level $l$} of a sequence of partitions if there is another sequence of partitions, compatible with the first one, in which the first $l$ partitions are the same. 
Observe that two vertices that are equivalent at level $l$ are equivalent at all levels $k, k<l$. Two vertices equivalent at some level belong to the same \emph{orbit}.


\section{Description of the Algorithm \emph{conauto-2.0}}
\label{s-algorithm}

Algorithm \emph{conauto-2.0} compares two graphs $G$ and $H$ for isomorphism. It works in the following way. First it computes a sequence of partitions for graph $G$. Then, unlike previous versions of conauto which only perform a partial (polynomial time) search for automorphisms, \emph{conauto-2.0} performs a full search for automorphisms. This yields a complete set of generators for the automorphism group of the graph.  The automorphisms found allow removing backtracking points from the sequence of partitions (a backtracking point at a level $l$ is removed, when all the vertices of the pivot cell in level $l$ are equivalent at that level). If, after removing backtracking points, the resulting sequence of partitions of $G$ still has backtracking points, a sequence of partitions for graph $H$ is generated, and the search for automorphisms and backtracking point removal in that graph is performed. Note that, if the sequence of partitions of $G$ does not have backtracking points, then it is not necessary to generate the sequence of partitions for $H$. 

Then, a backtracking process is used to find a match between the vertices of the two graphs. In this process, the sequence of partitions of one of the graphs is used as target, and an attempt is made to find a compatible sequence of partitions for the other graph. This uses the knowledge of the automorphism groups of the graphs, what drastically prunes the search space. The sequence of partitions with less backtracking points is chosen as the target for the match. Observe that if the target sequence of partitions has no backtracking points, the matching process has no backtracking.

\paragraph{Pivot cell selection.}
When the sequence of partitions of a graph is computed, at each backtracking point, a pivot cell must be chosen for vertex individualization. In conauto-2.0, at a level $k$, for each possible combination of cell size and available degree, the first cell in the partition, with that size and  available degree, is considered.  
The first vertex in each of these cells is individualized and the obtained partition is subsequently refined until an equitable partition is reached\footnote{Partition $\mathcal{S}=(S_1,...,S_r)$ is {\em equitable} if, for all $i,j \in \{1,...,r\}$ and all $u,v \in S_i$, the available degree of $u$ and $v$ with $S_j$ is the same.}.
For each vertex $x$, assume the equitable partition is reached at level $l_x$. If the partition at that level $l_x$ is a sub-partition of the partition at level $k$, then the cell of $x$ is chosen without testing any other. Otherwise, we sum the number of discarded vertices 
and the number of cells of this resulting partition. Then, the cell of the vertex $x$ which yields the biggest such sum is chosen as the pivot cell. Note that, since only one vertex in each cell is considered, the choice is not isomorphism invariant and, hence, cannot be used for canonical labeling. However, it works very well in practice for computing the automorphism group of a graph.

\paragraph{Recorded failures.}
In \cite{jllopez2009} one of the authors of this work suggested that recording failures during the search for automorphisms could help pruning the search space. Recently, the same idea has been proposed by Junttila and Kaski in \cite{springerlink:10.1007/978-3-642-19754-3_16}. The use of failures in conauto (version 2.0) and bliss (version 0.65 and above) are similar, but in our case they are used both in the search for automorphisms and during the matching process.

\paragraph{Search for automorphisms.}
The search for automorphisms is performed iteratively traversing all the backtracking points of the sequence of partitions from the last to the first. At each backtracking point, every vertex of the pivot cell (except the pivot vertex $p$ used in the original sequence of partitions), which has not been found yet to be equivalent to $p$, is chosen for vertex individualization. It is used to generate an alternative sequence of partitions using the same pivot cells and refinement procedures used in the original sequence of partitions. If a compatible sequence of partitions is found, the corresponding automorphism is recorded as a generator, and the vertex equivalences are updated accordingly. If all the vertices of the pivot cell at some level are found to be equivalent, then the backtracking point is removed.

The search for a compatible sequence of partitions that induces an automorphism of the graph is performed by a backtracking process that explores every feasible path in the search tree (all feasible pivot vertices at each backtracking point traversed). Known automorphisms and sub-partitions (as described below) are used to prune this search.

\paragraph{Use of sub-partitions to prune the search for automorphisms.} Consider the following definition.
\begin{definition}
\label{subpartition-def}
Let $k$ and $l$, $k<l$, be two backtracking levels. We say that $\mathcal{S}^l$ is a \emph{sub-partition} of $\mathcal{S}^k$ if $\not\exists i,j,h$ such that $i\ne j$, $S^l_i \subseteq S^k_h$, and $S^l_j \subseteq S^k_h$. (I.e., each cell of $\mathcal{S}^l$ is included in a different cell of $\mathcal{S}^k$.)
\end{definition}
As it will be proved in next section, when the search for automorphisms is being performed at level $k$, it is enough to consider partitions up to level $l$, where $l$ is the first level reached such that $\mathcal{S}^l$ is a sub-partition of $\mathcal{S}^k$. If a compatible alternative sequence of partitions can be generated up to level $l$, then it is possible to infer an automorphism without the need to generate any further partition. Observe that the property that partition $\mathcal{S}^l$ is a sub-partition of partition $\mathcal{S}^k$ can be evaluated before the search for automorphisms starts.

\section{Sub-partitions Theorem and Correctness}
\label{s-correctness}

In this section we present the main result that has been used to improve the performance of conauto-2.0. It is expressed as Theorem~\ref{contained-cells} below. 

Let $\mathsf{S}=(\mathcal{S}^0,...,\mathcal{S}^t)$ be a sequence of partitions for graph $G=(V,R)$, where $\mathcal{S}^i=(S^i_1,...,S^i_{r_i})$ and 
$V^i$ denotes the the set of vertices in partition
$\mathcal{S}^i$, for all $i \in \{0,...,t\}$. We consider two backtracking levels $k$ and $l$ so that $\mathcal{S}^l$ is a \emph{sub-partition} of $\mathcal{S}^k$ (see 
Definition \ref{subpartition-def}).
Let $p$
be the pivot vertex used for the vertex refinement at level $k$.
Consider a vertex 
$q \ne p$ (of the same cell) that, if used instead of $p$ as pivot vertex at level $k$,
generates an alternative sequence of partitions $\mathcal{T}^{k+1}, ..., \mathcal{T}^l$ that is compatible
with $\mathcal{S}^{k+1}, ..., \mathcal{S}^l$. We use $W^l$ to denote the set of vertices in partition
$\mathcal{T}^l$.

We define now the following sets. 
Let $E=V^k \setminus V^l$ and $E'=V^k \setminus W^l$
be the vertices discarded in the original and alternative sequences of partitions from level $k$ to level $l$, respectively.
Then, let us define $A=E\cap E'$ be the common vertices discarded,
$B=E \setminus A$ the vertices discarded only in the original (sub)sequence,
$C=E' \setminus A$ the vertices discarded only in the alternative (sub)sequence,
and $D=V^l \cap W^l$ the common vertices not discarded.
Clearly, $E=A\cup B$, and $E'=A\cup C$. 
Let us also, for each $X \in \{E,E',A,B,C,D\}$, define the subset $X_i=X \cap S^k_i$.
Observe that $|E_i|=|E'_i|$, and hence $|B_i|=|C_i|$.

\begin{Algorithm}{
\caption{Map the vertices of $V^l$ with vertices of $W^l$}
\label{fillperm}
\begin{footnotesize}
\begin{algo}{$\mathit{ComputeSubPartition}(G,S,T,V^l,W^l):\mathrm{void}$}
\> {\bf for} each $v_i \in V^l$ {\bf do} $w_i \leftarrow v_i$ {\bf end for} \nl
\> {\bf for all} $e'_i \in C$ {\bf do} \nl
\> \> $j \leftarrow i$ \nl
\> \> {\bf while} $e_j \in A$ {\bf do} $j \leftarrow k : e_j = e'_k$ {\bf end while} \nl
\> \> $w_l \leftarrow e_j$ where $l : w_l = e'_i$ \nl
\> {\bf end for}
\end{algo}
\end{footnotesize}
\BB\BB
}
\end{Algorithm}

As it was observed in \cite{jllopez2009}, the graph induced by $E$ is isomorphic to the graph induced by $E'$, and there is an isomorphism of them that matches the vertices
in $E_i$ to those in $E'_i$, for all $i$. Furthermore, let $e_1, e_2, ..., e_{|E|}$ and $e'_1, e'_2, ..., e'_{|E|}$ be the vertices in $E$ and $E'$, respectively, in the order in which they have been removed from their respective sequences of partitions. Then, mapping $e_i$ to $e'_i$ 
gives such isomorphism.
The (sub)sequence of partitions $\mathcal{S}^l, ..., \mathcal{S}^t$ also imposes an order among the vertices of $V^l$. Let $v_1, v_2, ..., v_{|V^l|}$ be such order.
Algorithm~\ref{fillperm} defines an order $w_1, w_2, ..., w_{|V^l|}$ of the vertices of $W^l$ so that the mapping of $v_i$ to $w_i$, for each $v_i \in V^l$, extends the isomorphism induced by the sequences of partitions to an automorphism of graph $G$. 
What Algorithm~\ref{fillperm} does to construct the order $w_1, w_2, ..., w_{|V^l|}$ is to start as $v_1, v_2, ..., v_{|V^l|}$ and replace each vertex in this sequence that belongs to $C$ by a different vertex from $B$. 

\begin{theorem}
\label{contained-cells}
Let $\mathcal{T}^l$ and $\mathcal{S}^l$ be two subpartitions of $\mathcal{S}^k$, compatible between them, and $k < l$, the identity mapping for vertices in $V \setminus V^k$, combined with the mapping between $E$ and $E'$ induced by the sequence of partitions, and the mapping between $V^l$ and $W^l$ obtained with Algorithm~\ref{fillperm},
define an automorphism of graph $G$. 
\end{theorem}

\section{Performance Evaluation}
\label{performance}
In this section we compare the performance of conauto-2.0 against other algorithms for graph isomorphism testing and automorphism group computation. The experiments have been carried out in an Intel i7 Q 720 @ 1.6GHz with 8GiB of RAM under Ubuntu 10.04. All programs have been compiled with gcc 4.4.3 with their respective default configuration, but modified to perform isomorphism testings or automorphism group computation, depending on the experiment. The first experiment considers some graph families, and is intended to see how the size of the graphs affect the running time of isomorphism testing programs. Each point shown in the plots corresponds to the average running time of 100 executions with different instances of the corresponding graph. In the second experiment, some singular cases have been considered, to compare the running times for automorphism group computation. The times shown are for a single instance of each graph. The CPU time limit of each execution has been set to 10,000 seconds for both experiments.

\begin{figure*}[t!]
\centering
\includegraphics{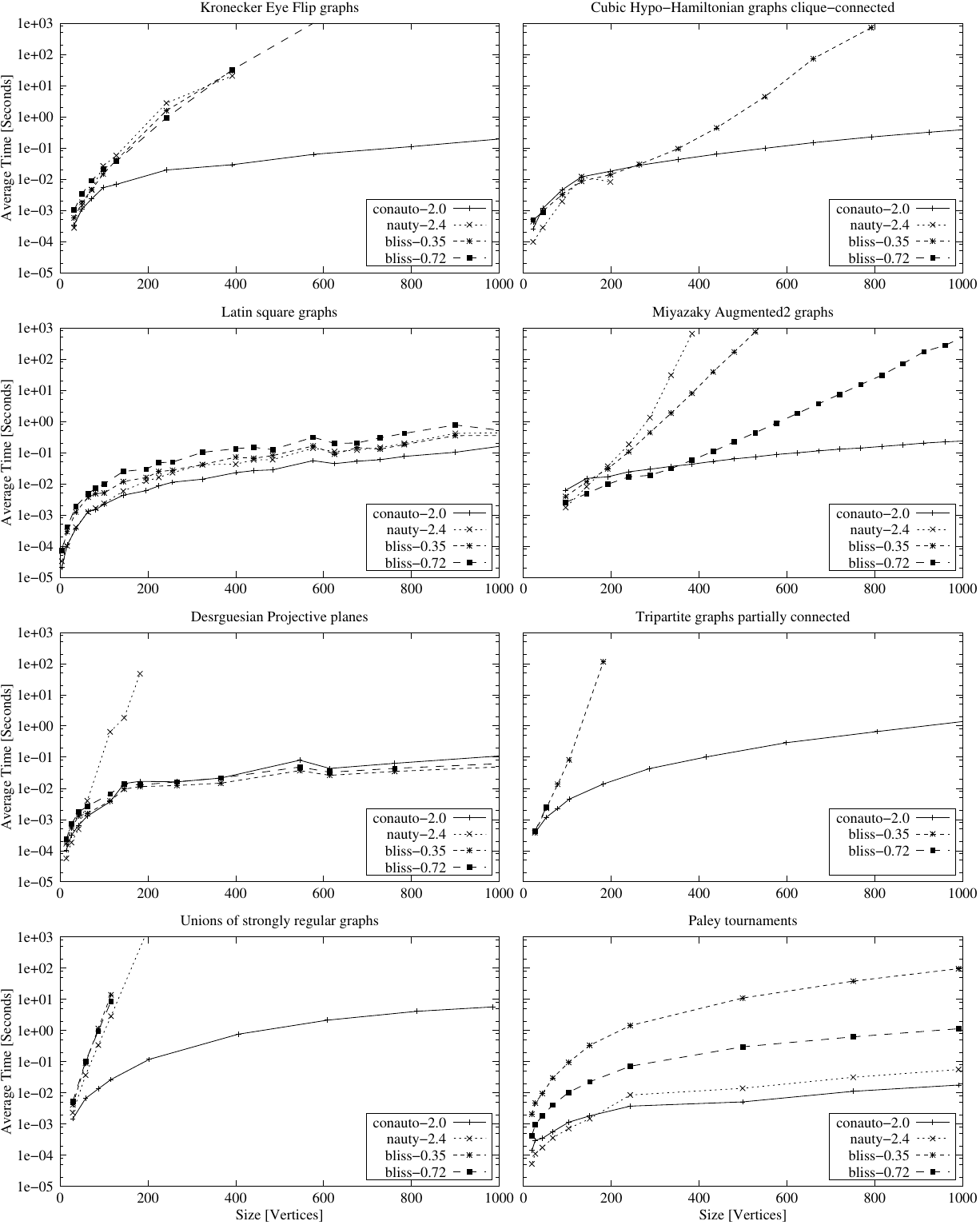}
\caption{Performance evaluation for testing isomorphism.}
\label{per-fig}
\end{figure*}

Many graph families are used by the different graph isomorphism and canonical labeling algorithms developers. Unfortunately, there is not enough room here for all of them, and not all of them are significant. In Figure~\ref{per-fig} we show the results we have obtained with a few selected families. They are of different types, dense and sparse, they are all hard for most algorithms. The families considered are the following.

\begin{packed_itemize}
\item
Kronecker Eye Flip graphs (KEF).
They are part of the benchmark of bliss \cite{bliss-benchmark}.
\item
Cubic Hypo-Hamiltonian graphs clique-connected ($CCH\_cc$).
This family is proposed by the authors. It is built using as basic block two non-isomorphic Hypo-Hamiltonian graphs with 22 vertices. Both graphs have four orbits of sizes: one, three, six, and twelve. A graph $CHH\_cc(m,n)$ has $n$ \emph{complex} commponents built from $m$ basic components. The components of a complex component are connected through a complete $m$-partite graph using the graphs that belong to the orbits of size three of each basic component. The $n$ complex components are interconnected with a complete $n$-partite graph using the vertices of each complex component that belong to the orbits of size one in the basic components. With this family, bliss-0.72 returns an internal error for many benchmark graphs of 88 or more vertices.
\item
Latin square graphs.
These are $L_3(n)$ latin squares of order $n$, for non-prime $n$. They were hard for the previous versions of conauto, but not for conauto-2.0.
\item
Miyazaki Augmented2 graphs.
These are taken from the benchmark of bliss \cite{bliss-benchmark}. For this family of graphs, bliss-0.72 gives a significant improvement over nauty-2.4 and bliss-0.35, but its performace is far from that of conauto-2.0 for large instances.
\item
Point-line graphs of Desarguesian projective planes.
This family was extremely hard for previous versions of conauto. However, thanks to the adaptive selection of the pivot cells, it has now a performance comparable with both versions of bliss.
\item
Tripartite graphs, partially connected.
These graphs are proposed by the authors, and are built from two $13$ vertices directed tripartite graphs. Four vertices of each component have an arc to three vertices of each other component. Thus, each component is connected to all the other components of the graph. For graphs of 78 vertices or more, bliss-0.72 generates internal errors in many cases. There are not results for nauty because for some instances on 26 vertices, nauty-2.4 is not able to finish within the time limit.
\item
Unions of strongly regular graphs.
The components are strongly regular graphs of 29 vertices each. Each vertex in one component is connected to every vertex of the other components. Thus, this family is extremely dense. Observe that only conauto-2.0 is able to finish within the time limit for graphs of more than 203 vertices.
\item
Paley tournaments. For this family of graphs all algorithms exhibit nice time complexity. However, conauto-2.0 is clearly the fastest of the four programs tested.
\end{packed_itemize}
To conclude, we can say that conauto-2.0 outperforms all the other algorithms or has a similar behavior for the graph families considered.

\begin{figure*}[t!]
\centering
\begin{footnotesize}
\begin{tabular}{ | c | c | r | r | r | r | r | r | }
  \hline                       
  Graph 	& (\#nodes,  & bliss0.35 & bliss0.72 & nauty2.4 & saucy2.1 & Traces & conauto2.0 \\
   & \#edges) & \emph{time(s)} & \emph{time(s)} & \emph{time(s)} & \emph{time(s)} & \emph{time(s)} & \emph{time(s)} \\
  \hline                       
  \hline                       
  had-236 	& (944,111864) 	& TL 	  & 130.939 & TL      & TL 	& 137.135 & 1704.883 \\
  had-sw-112 	& (448,25312) 	& 829.983 & 9.716   & TL      & 8256.973& 8.950   & 13.696 \\
  pp16-6	& (546,4661)	& TL	  & 2449.465& TL      &	TL	& 1.034	  & 162.722 \\
  pp16-8 	& (546,4641) 	& 19.145  & 4.506   & TL      & TL	& 0.219   & 0.454 \\
  pp16-9 	& (546,4641) 	& 650.374 & 48.569  & TL      & TL	& 0.277   & 5.339 \\
  mz-aug-50	& (1000,2300)	& 0.156	  & 0.080   & TL      & TL	& 19.271  & 0.110 \\
  k-100		& (100,4950)	& 0.042	  & 0.079   & 0.015   & 0.000   & 29.104  & 0.002 \\
  latin-sw-30-11& (900,39150)	& 41.699  & 1.288   & 41.776  & 22.718	& 18.342  & 2.788 \\
  s3-3-3-3 	& (11076,20218) & 0.086   & 0.029   & 12.231  & 0.009   & 2127.104& 1.618 \\
  rnd-3-reg-3K 	& (3000,4500) 	& 0.326   & 0.200   & 283.349 & 0.068 	& 0.339   & 3.253 \\
  rnd-3-reg-10K & (10000,15000) & 4.228   & 2.406   & TL      & 0.451   & 4.530  & 68.474 \\
  \hline  
\end{tabular}
\end{footnotesize}
\caption{Performance evaluation for computing automorphism groups. TL means that the execution has exceeded the time limit.}
\label{aut-fig}
\end{figure*}

Figure~\ref{aut-fig} shows the results of the second experiment. As mentioned, in this experiment we test the performance of computing the automorphism group of different graphs. All the graphs used have been taken from the benchmark of bliss \cite{bliss-benchmark}. As can be seen, the best performance is presented by bliss0.72, Traces, and conauto-2.0, without a clear
winner among them.



\begin{thebibliography}{10}

\bibitem{ConteFSV03}
Donatello Conte, Pasquale Foggia, Carlo Sansone, and Mario Vento.
\newblock Graph matching applications in pattern recognition and image
  processing.
\newblock In {\em IEEE International Conference on Image Processing}, volume~2,
  pages 21--24, Barcelona, Spain, September 2003.

\bibitem{DBLP:conf/dac/DargaLSM04}
Paul~T. Darga, Mark~H. Liffiton, Karem~A. Sakallah, and Igor~L. Markov.
\newblock Exploiting structure in symmetry detection for cnf.
\newblock
{\em DAC},
  pages 530--534. ACM, 2004.

\bibitem{Faulon98}
Jean-Loup Faulon.
\newblock Isomorphism, automorphism partitioning, and canonical labeling can be
  solved in polynomial--time for molecular graphs.
\newblock {\em Journal of chemical information and computer science},
  38:432--444, 1998.

\bibitem{bliss-benchmark}
Tommi Junttila.
\newblock Benchmark graphs for evaluating graph automorphism and canonical
  labeling algorithms.
\newblock Laboratory for Theoretical Computer Science, Helsinki University of
  Technology, 2009.
\newblock {http://www.tcs.hut.fi/Software/bliss/benchmarks/index.shtml}.

\bibitem{springerlink:10.1007/978-3-642-19754-3_16}
Tommi Junttila and Petteri Kaski.
\newblock Conflict propagation and component recursion for canonical labeling.
\newblock
{\em
  Theory and Practice of Algorithms in (Computer) Systems},
 {\em Lecture Notes in Computer Science}, 6595:151--162. Springer Berlin /
  Heidelberg, 2011.

\bibitem{DBLP:conf/alenex/JunttilaK07}
Tommi~A. Junttila and Petteri Kaski.
\newblock Engineering an efficient canonical labeling tool for large and sparse
  graphs.
\newblock In {\em ALENEX}. SIAM, 2007.

\bibitem{DBLP:conf/sat/KatebiSM10}
Hadi Katebi, Karem~A. Sakallah, and Igor~L. Markov.
\newblock Symmetry and satisfiability: An update.
\newblock
{\em SAT},
  of {\em Lecture Notes in Computer Science}, 6175:113--127. Springer, 2010.

\bibitem{DBLP:conf/wea/Lopez-PresaA09}
Jos{\'e}~Luis L{\'o}pez-Presa and Antonio~Fern{\'a}ndez Anta.
\newblock Fast algorithm for graph isomorphism testing.
\newblock
{\em SEA},
 {\em Lecture Notes in Computer Science}, 5526:221--232. Springer, 2009.

\bibitem{McK81}
Brendan~D. McKay.
\newblock Practical graph isomorphism.
\newblock {\em Congressus Numerantium}, 30:45--87, 1981.

\bibitem{nautyP}
Brendan~D. McKay.
\newblock The nauty page.
\newblock Computer Science Department, Australian National University, 2010.
\newblock {http://cs.anu.edu.au/$\sim$bdm/nauty/}.

\bibitem{Miyazaki97}
Takunari Miyazaki.
\newblock The complexity of {McKay's} canonical labeling algorithm.
\newblock
{\em Groups and
  Computation II}, volume~28 of {\em DIMACS Series in Discrete Mathematics and
  Theoretical Computer Science}, pages 239--256. American Mathematical Society,
 1997.

\bibitem{DBLP:journals/corr/abs-0804-4881}
Adolfo Piperno.
\newblock Search space contraction in canonical labeling of graphs (preliminary
  version).
\newblock {\em CoRR}, abs/0804.4881, 2008.

\bibitem{jllopez2009}
Jos\'e Luis~L\'opez Presa.
\newblock {\em Efficient Algorithms for Graph Isomorphism Testing}.
\newblock Doctoral thesis, Escuela T\'ecnica Superior de Ingenier\'{\i}a de
  Telecomunicaci\'on, Universidad Rey Juan Carlos, Madrid, Spain, March 2009.
\newblock Available at http://www.diatel.upm.es/jllopez/tesis/thesis.pdf.

\bibitem{Tener2009}
G.~Tener.
\newblock {\em Attacks on difficult instances of graph isomorphism: sequential
  and parallel algorithms}.
\newblock Phd thesis, University of Central Florida, 2009.

\bibitem{Tener_Deo_2008}
Greg Tener and Narsingh Deo.
\newblock Attacks on hard instances of graph isomorphism.
\newblock {\em Journal of Combinatorial Mathematics and Combinatorial
  Computing}, 64:203--226, 2008.

\bibitem{TinKlin99}
Gottfried Tinhofer and Mikhail Klin.
\newblock Algebraic combinatorics in mathematical chemistry. {Methods} and
  algorithms {III}. {Graph} invariants and stabilization methods.
\newblock Technical Report TUM-M9902, Technische Universit\"at M\"unchen, March
  1999.

\end{thebibliography}

\newpage

\appendix
\section{Proof of Theorem \ref{contained-cells}}
The proof of following lemmas can be found in \cite[chapter 7, p. 70-75]{jllopez2009}.

%
 
\begin{figure}[!h]
\begin{center}
\begin{tabular}{c|c|c|}
 \multicolumn{1}{c}{ } & \multicolumn{1}{c}{$E'_1$} & \multicolumn{1}{c}{$T^l_1$} \\
\cline{2-3} \multicolumn{1}{c|}{$E_1$} & $A_1$ & $B_1$ \\
\cline{2-3} \multicolumn{1}{c|}{$S^l_1$} & $C_1$ & $D_1$ \\
\cline{2-3}
\end{tabular}
\hspace{1cm}
{\large\ldots}
\hspace{0,7cm}
\begin{tabular}{c|c|c|}
 \multicolumn{1}{c}{ } & \multicolumn{1}{c}{$E'_r$} & \multicolumn{1}{c}{$T^l_r$} \\
\cline{2-3} \multicolumn{1}{c|}{$E_r$} & $A_r$ & $B_r$ \\
\cline{2-3} \multicolumn{1}{c|}{$S^l_r$} & $C_r$ & $D_r$ \\
\cline{2-3}
\end{tabular}
\end{center}
\label{division-S-k-i}
\caption{Partition of $S^k_i$ into subsets $A_i$, $B_i$, $C_i$, and $D_i$, for all $i$.}
\end{figure}


\begin{lemma}
\label{uno-a-todos}
Let $M=\mathit{Adj}(G)$. For each $u \in E$, for all $i\in\{1,...,r_l\}$, for all $v,w \in S^l_i$, $M_{uv} = M_{uw}$ and $M_{vu} = M_{wu}$. Similarly,
For each $u' \in E'$, for all $i\in\{1,...,r_l\}$, for all $v',w' \in T^l_i$, $M_{u'v'} = M_{u'w'}$ and $M_{v'u'} = M_{w'u'}$.
\end{lemma}

\begin{lemma}
\label{Bi-Cj-todos-a-todos}
For each $i,j\in\{1,...,r_l\}$, there is some $a\in\{0,...,3\}$ such that
for all $u\in B_i$, $v\in C_i$, $w\in D_i$, $u'\in B_j$, $v'\in C_j$, and $w'\in D_j$,
$M_{uv'}=M_{uw'}=M_{vu'}=M_{vw'}=M_{wu'}=M_{wv'}=a$ and $M_{u'v}=M_{u'w}=M_{v'u}=M_{v'w}=M_{w'u}=M_{w'v}=a^{-1}$. 
(We define $0^{-1}=0, 1^{-1}=2, 2^{-1}=1$, and $3^{-1}=3$.)
\end{lemma}

Let us define two families of partitions of $A_i$ for $i,j\in\{1,...,r\}$:
$$A_i^{cj}=\{x\in A_i : \forall u \in B_i, v' \in C_j, M_{xv'} = M_{uv'} \}$$
$$A_i^{nj}=\{x\in A_i : \forall u \in B_i, v' \in C_j, M_{xv'} \ne M_{uv'} \}$$
Note that, since the vertices of $A_i$ are unable to distinguish among the vertices
of $C_j$, then, if $M_{xv'} \ne M_{uv'}$ for some $u \in B_i$ or some $v' \in C_j$,
then $M_{xv'} \ne M_{uv'}$ for all $u \in B_i$ and all $v' \in C_j$. Hence, each pair
of sets $A_i^{cj}$ and $A_i^{nj}$ defines a partition of $A_i$. Note also that, since
each vertex in $A_i$ has the same type of adjacency with all the vertices in
$B_i \cup C_i \cup D_i$ (from Lemma~\ref{uno-a-todos}), then for all $x \in A_i^{cj}$,
$u\in B_i$, $v\in C_i$, $w\in D_i$, $u'\in B_j$, $v'\in C_j$, and $w'\in D_j$,
$M_{xu'}=M_{xv'}=M_{xw'}=M_{uv'}=M_{uw'}=M_{vu'}=M_{vw'}=M_{wu'}=M_{wv'}$ (from
Lemma~\ref{Bi-Cj-todos-a-todos}).

\begin{figure}[!h]
\begin{center}
\begin{tabular}{c|c|c|}
\multicolumn{1}{c}{ } & \multicolumn{1}{c}{$E'_i$} & \multicolumn{1}{c}{$T^l_i$} \\
\cline{2-3} \multicolumn{1}{c|}{$E_i$} & \slashbox{$A^n_i$}{$A_i^c$} & $B_i$ \\
\cline{2-3} \multicolumn{1}{c|}{$S^l_i$} & $C_i$ & $D_i$ \\
\cline{2-3}
\end{tabular}
\end{center}
\label{division-A-i}
\caption{Partition of $A_i$ into subsets $A^c_i$, and $A^n_i$.}
\end{figure}

\begin{lemma}
\label{An-themselves}
For all $i\in\{1,...,r\}$, let $A_i^c=\bigcap_{j=1}^r A_i^{cj}$, and let
$A_i^n=\bigcup_{j=1}^r A_i^{nj}$. Then, any isomorphism of $G_E$ and $G_{E'}$
that maps $G_{E_i}$ to $G_{E'_i}$, maps the vertices in $A_i^n$ among themselves.
\end{lemma}

\begin{lemma}
\label{Gb-iso-Gc}
$G_B$ is isomorphic to $G_C$, and there is an isomorphism of them that matches the vertices
in $B_i$ to those in $C_i$, for all $i\in\{1,...,r\}$.
\end{lemma}

\begin{lemma}
\label{GVl-iso-GWl}
$G_{V^l}$ and $G_{W^l}$ are isomorphic, and there is an isomorphism of them that maps the vertices
in $S^l_i$ to the vertices of $T^l_i$ for all $i \in \{1,...,r_l\}$.
\end{lemma}

Let $l$ be a backtracking level and let $\mathcal{S}^l=(S^l_1,...,S^l_r)$ be the partition at that level, then observe that, for all $i \in \{1,...,r\}$, $G_{S^l_i}$ is regular.

Once we reach level $l$, generating an alternative sequence of partitions, which is subpartition of a level $k < l$, we apply Algorithm \ref{fillperm} due to obtain an automorphism of $G$.



\begin{observation}
\label{o-one-all}
A vertex discarded up to level $l$ has the same adjacency with all the vertices of each cell of level $l$. Otherwise it would have split such cell.
\end{observation}

\begin{observation}
\label{o-same-cell}
The vertices that belong to the same cell at level $k$ and have not been discarded up to level $l$, must belong to the same cell at level $l$, since the partition of level $l$ is a subpartition of the partition of level $k$.
\end{observation}

\begin{observation}
\label{eq-inside-cell}
For all $i\in \{1,...,|E|\}$, $e_i$ and $e'_i$ belong to the same cell in level $k$. This follows from the compatibility of the sequences of partitions.
\end{observation}

\begin{observation}
\label{B-C-same-cell}
Algorithm~\ref{fillperm} substitutes a vertex that belongs to a cell in level $k$ with a vertex that belonged to the same cell at that level $k$. This follows from Observation~\ref{eq-inside-cell}.
\end{observation}

\begin{proofof}{Theorem~\ref{contained-cells}}

Let $x_1,x_2,...,x_{|V\setminus V^k|}$ be the order induced by the sequences of partitions on the vertices discarded up to level $k$.
Then $x_1,...,x_{|V\setminus V^k|},e_1,...,e_{|E|},v_1,...,v_{|V^l|}$ is the order induced by the sequence of partitions $\mathsf{S}$. Let us call this order \emph{base}.
Analogously, $x_1,...,x_{|V\setminus V^k|},e'_1,...,e'_{|E|},w_1,...,w_{|V^l|}$ is the new \emph{generator} computed. We now prove that they define an automorphism of the graphs, i.e.:
\begin{packed_enum}
\item\label{t-case1}
For each $i,j\in \{1,...,|V\setminus V^k|\}, M_{x_i x_j}=M_{x_i x_j}$. (Trivial).
\item\label{t-case2}
For each $i,j\in \{1,...,|E|\}, M_{e_i e_j}=M_{e'_i e'_j}$.
\item\label{t-case3}
For each $i,j\in \{1,...,|V^l|\}, M_{v_i v_j}=M_{w_i w_j}$.
\item\label{t-case4}
For each $i\in \{1,...,|V\setminus V^k|\}, j\in\{1,...,|E|\}, M_{x_i e_j}=M_{x_i e'_j}$.
\item\label{t-case5}
For each $i\in \{1,...,|V\setminus V^k|\}, j\in\{1,...,|V^l|\}, M_{x_i v_j}=M_{x_i w_j}$.
\item\label{t-case6}
For each $i\in \{1,...,|E|\}, j\in\{1,...,|V^l|\}, M_{e_i v_j}=M_{e'_i w_j}$.
\end{packed_enum}

Case~\ref{t-case2} follows directly from the compatibility of the sequences of partitions.

Case~\ref{t-case4} follows directly from the compatibility of the sequences of partitions.

Case~\ref{t-case5} follows from the fact that, if no discarded vertex of a sequence of partitions has split a cell, that is because that vertex had the same adjacency with all the vertices of this cell. Since the partitions at level $l$ are subpartitions of the partition at level $k$ and Observation~\ref{B-C-same-cell}, all the vertices discarded before level $k$ have the same adjacencies with all the vertices of each cell of level $l$. Otherwise, those cells would have been split before level $k$.

From Observation~\ref{B-C-same-cell}, the vertices of a cell of level $l$ that belonged to come cell of level $k$ are substituted by Algorithm~\ref{fillperm} by vertices that also belonged to that cell of level $k$. Hence, in level $l$ they belonged to corresponding cells. Hence, from this fact, Lemma~\ref{uno-a-todos} and the compatibility of the sequences of partitions, Case~\ref{t-case6} follows.

From Lemma~\ref{GVl-iso-GWl}, there is an isomorphism that maps the vertices in corresponding cells of level $l$. It is easy to see that mapping the vertices of $D$ to themselves, we obtain one automorphism of $G_D$ (which maps the vertices of $D_i$ among themselves for all $i$). We only need to prove now that having substituted the vertices of $C_i$ by those of $B_i$ preserves the adjacencies among them, and from them to the vertices of $D$.

From Lemma~\ref{Bi-Cj-todos-a-todos}, the vertices of $B_i$ have with the vertices of $D_j$ the same adjacencies that have the vertices of $C_i$ with those of $D_j$, for all $i,j$. Hence, the substitution preserves these adjacencies.

Finally we need to prove that Algorithm~\ref{fillperm} preserves the adjacencies among the vertices of $B$ and the vertices of $C$. 

From Lemma~\ref{An-themselves}, we know that the vertices of $B$ and $C$ are not mapped to any vertex of $A^n$. Consider first the vertices of $B$ that are mapped by the isomorphism induced by the sequences of partitions to vertices of $C$. Each of these vertices of $B$ have the same type of adjacency with the other vertices of $B$ than the corresponding vertex of $C$ with the other vertices of $C$.

Let us consider now the vertices of $B$ and $C$ that are mapped to vertices of $A$ (in fact, from Lemma~\ref{An-themselves}, we know that they are mapped to vertices of $A^c$). In this case, Algorithm~\ref{fillperm} maps each vertex $b_i \in B_i$ to a vertex $c_i\in C_i$ such that the isomorphism induced by the sequences of partitions maps $b_i$ to some vertex $a_i$ in $A^c_i$, and this vertex $a_i$ is mapped to $c_i$ (there could be a chain of vertices in $A^c_i$ instead of only one, but that makes no difference in the argument). Since the vertices of $A^c_i$ have the same type of adjacency with all the vertices of $B_j,C_j$, for all $j$ (from the definition of $A^c_i$). Hence, if vertex $b_i$ was mapped to $a_i$, then it had the same type of adjacency with all the vertices of $B_j$, Since $a_i$ was mapped to $c_i$, then $c_i$ also had the same type of adjacency with all the vertices of $C_j$, (the same that $a_i$ had). Hence, vertex $c_i$ had the same type of adjacency with all the vertices of $C_j$, than vertex $b_i$ had with all the vertices of $B_j$, for all $j$. Thus we prove that the mapping generated by Algorithm~\ref{fillperm} yields an isomorphism of $G_B$ and $G_C$, completing the proof.

\end{proofof}

\end{document}